\documentclass[conference]{IEEEtran}
\IEEEoverridecommandlockouts

\ifCLASSINFOpdf

\else

\fi

\usepackage{calc}
\usepackage{graphics,color,subfigure,wrapfig}

\usepackage{url}

\usepackage[breaklinks]{hyperref}
\usepackage{breakurl}

\usepackage{color}
\usepackage{amsthm}
\usepackage{amsmath,amssymb}
\usepackage{amsfonts}
\usepackage{array}
\newcolumntype{P}[1]{>{\centering\arraybackslash}p{#1}}
\usepackage[tmargin=0.55in,bmargin=0.55in,lmargin=0.625in,rmargin=0.625in]{geometry}
\usepackage{multirow}
\usepackage{graphicx}
\usepackage{standalone}
\usepackage{booktabs}
\usepackage{algpseudocode}
\usepackage{algorithm2e}
\let\oldnl\nl
\newcommand{\nonl}{\renewcommand{\nl}{\let\nl\oldnl}}
\usepackage{makecell}
\usepackage[bottom]{footmisc}
\usepackage{enumitem}
\theoremstyle{definition}
\usepackage{tikz}
\usepackage{xcolor}

\makeatletter
\renewcommand{\@algocf@capt@plain}{above}
\makeatother
\RestyleAlgo{ruled}
\LinesNumbered

\newtheorem{lemma}{Lemma}

\newtheorem{definition}{Definition}

\newtheorem{example}{Example}

{\begin{list}{}%
         {\setlength{\leftmargin}{#1}}%
         \item[]%
}
{\end{list}}

\newcounter{relctr} 
\everydisplay\expandafter{\the\everydisplay\setcounter{relctr}{0}} 

\newcommand\labelrel[2]{%
  \begingroup
    \refstepcounter{relctr}%
    \stackrel{\textnormal{(\alph{relctr})}}{\mathstrut{#1}}%
    \originallabel{#2}%
  \endgroup
}
\AtBeginDocument{\let\originallabel\label} 

\begin{document}

\title{CAMR: Coded Aggregated MapReduce}

\author{\IEEEauthorblockN{Konstantinos Konstantinidis and Aditya Ramamoorthy}
\IEEEauthorblockA{Department of Electrical and Computer Engineering\\
Iowa State University\\
Ames, IA 50010\\
Email: \{kostas, adityar\}@iastate.edu \thanks{This work was supported in part by the National Science Foundation (NSF) under grant CCF-1718470.}}
}

\maketitle

\begin{abstract}

Many big data algorithms executed on MapReduce-like systems have a shuffle phase that often dominates the overall job execution time. Recent work has demonstrated schemes where the communication load in the shuffle phase can be traded off for the computation load in the map phase. In this work, we focus on a class of distributed algorithms, broadly used in deep learning, where intermediate computations of the same task can be combined. Even though prior techniques reduce the communication load significantly, they require a number of jobs that grows exponentially in the system parameters. 
This limitation is crucial and may diminish the load gains as the algorithm scales. We propose a new scheme which achieves the same load as the state-of-the-art while ensuring that the number of jobs as well as the number of subfiles that the data set needs to be split into remain small. 

\end{abstract}

%

\section{Introduction}

The recent growth of big data analytics whereby a large amount of data on the orders of petabytes or more needs to be processed in a fast manner has fueled the development of several distributed programming models running on clusters of commodity servers. 
Some characteristic examples are MapReduce \cite{DeanG08},
Hadoop \cite{HADOOP} 
and Spark \cite{apache_spark}.

In these frameworks, the data set is split into disjoint subfiles stored across the worker nodes. The computation takes place in three steps. 
Initially, the processing servers \emph{map} the input subfiles to \emph{intermediate values} having the form of (key, value) pairs. In the next \emph{shuffle} step, the intermediate pairs are exchanged between the servers. In the final \emph{reduce} step, each server computes a set of output functions defined based on the keys. By virtue of their simplicity, scalability and fault-tolerance, these frameworks are becoming ubiquitous and have gained significant momentum within both industry and academia.
They are well suited for several applications including machine learning \cite{compressed_CDC}, \cite{he_resnet}, graph processing \cite{saurav_g_analytics}, data sorting \cite{konstantinidis_ramamoorthy_globecom}
and web logging \cite{DeanG08}.

Compelling evidence obtained on large scale clusters suggests that the time spent merely on communication often dominates the execution time. For example, by analyzing a week-long trace from  Facebook's Hadoop cluster, the authors of \cite{Chowdhury_etal11} demonstrated that {\it ``on average, $33$\% of the overall job execution time is spent on data shuffling."} Similar effects have been reported in the work of \cite{GuoRZ13} on other shuffle-heavy operations such as SelfJoin, TeraSort and RankedInvertedIndex which underlie many deep learning algorithms. Distributed graph analytics also suffer from long communication phases as observed in \cite{saurav_g_analytics}, 
accounting for up to $50$\% of the overall execution time in representative cases 
\cite{Chen_graph14}.

In this paper, we focus on distributed algorithms for which the intermediate values of a particular job computed during the Map phase can be combined locally by the servers before the transmission. 
This kind of computation is predominant in machine learning (e.g., ImageNet classification \cite{he_resnet} and stochastic gradient descent \cite{tandon_gradient}). Another use case would be the matrix-vector multiplications performed during the forward and backward propagation in neural networks (\emph{cf.} \cite{Dally_nips_tutorial}). In our context, computing each of these products constitutes a job. We could also consider training multiple models simultaneously, as long as they have the same dimensionality. This so-called \emph{compression technique} was initially investigated in \cite{DeanG08} by the means of a ``combiner function" which merges multiple intermediate values with the same key computed from different Map functions. 



The work of \cite{LiMA16} proposed an approach (inspired by coded caching) for trading off communication load with computation load in MapReduce-like systems. This was extended in \cite{compressed_CDC} to Compressed Coded Distributed Computing (CCDC), where compressible functions were considered.  In prior work, we addressed one limitation of \cite{LiMA16}, namely the requirement that jobs need to fit very finely to obtain the promised communication load. Our approach demonstrated a deep relationship between this problem and a class of combinatorial
structures called \emph{resolvable designs} while achieving significant speedup compared to the state-of-the-art.


\subsection{Main contributions of our work}

It turns out that \cite{compressed_CDC} has a limitation of a similar flavor. In this case the number of jobs needs to scale exponentially in the problem parameters to obtain the promised reduction in communication load.

In this work, we extend our algorithm to applications where intermediate values can be compressed and we substantially reduce the requirement on the number of jobs compared to prior literature. The immediate benefit that stems from this fact is that as the size of the cluster increases, the required number of MapReduce jobs (and hence the total number of subfiles) does not scale exponentially. The implicit benefit is that a low requirement on the number of jobs decreases the encoding complexity. This is important since, as we have shown in \cite{konstantinidis_ramamoorthy_globecom}, increasing the number of tasks scales the overhead of the encoding complexity and can diminish any gains in the communication load. We expect a similar type of phenomenon in the current setting. 

Our new scheme is named \emph{coded aggregated MapReduce} (abbreviated, \emph{CAMR}). We characterize the achievable communication load of CAMR and show that it matches the state-of-the-art.
The next section gives the general problem formulation, while in the remaining sections, we describe our scheme, analyze the achievable load and compare it with other combining methods. 


\section{Problem Formulation}
\label{sec:formulation}
Our goal is to process $J$ distributed computing jobs (denoted $\mathcal{J}_1,\dots,\mathcal{J}_J$) in parallel on a cluster of $K$ homogeneous servers $U_1,\dots,U_K$, i.e., machines that have similar computational power. The data set of each job is partitioned into $N$ disjoint and equal-sized subfiles. The subfiles of the $j$-th job are denoted by $n^{(j)},n=1,\dots,N$. A total of $Q$ output functions, denoted $\phi_q^{(j)},q=1,...,Q$, need to be computed for each job. 
Note that these $Q$  functions may be different across different jobs. We examine a special class of functions that possess the \emph{aggregation} property. 

\begin{definition}
In database systems, an \emph{aggregate function} $\phi$ is one that is both associative and commutative.
\end{definition}

For example, in jobs with linear aggregation the evaluation of each output function can be decomposed as the sum of $N$ \emph{intermediate values}, one for each subfile, i.e., for $q=1,\dots,Q$,
$$\phi_q^{(j)}(1^{(j)},\dots,N^{(j)})=\nu_{q,1}^{(j)}+\nu_{q,2}^{(j)}+\dots+\nu_{q,N}^{(j)},$$
where $\nu_{q,n}^{(j)}=\phi_q^{(j)}(n^{(j)})$ and each such value is assumed to be of size $B$ bits. In what follows we use $\alpha(\nu_{q,1}^{(j)},\dots,\nu_{q,m}^{(j)})$ to denote the aggregation of $m$ intermediate values $\nu_{q,1}^{(j)},\dots,\nu_{q,m}^{(j)}$ of the same function $\phi_q^{(j)}$ and job $\mathcal{J}_j$ into a single ``compressed" value.


A master node judiciously places each subfile on at least one server before initiating the algorithm. 

\begin{definition}
The storage fraction $\mu\in[1/K,1]$ of a distributed computation scheme is the fraction of the data sets across all jobs that each machine locally caches.
\end{definition}

Our formulation assumes that $K$ divides $Q$ so that each server is assigned to $Q/K$ functions per job.
However, our proposed algorithm and the main results can be obtained as a simple extension of the case when each server is computing one function. For example, one can repeat the Shuffle phase $Q/K$ times. Owing to this fact, we will only present the case of $Q=K$. 

The framework starts with the \emph{Map} phase during which the servers (in parallel)
``map" every subfile $n^{(j)}$ to the values $\{\nu_{1,n}^{(j)},\dots,\nu_{Q,n}^{(j)}\}$. 
Following this, the servers multicast the computed intermediate values amongst one another via a shared link in the \emph{Shuffle} phase. 
In the final \emph{Reduce} phase, server $k$ computes (or reduces) $\phi_k^{(j)}(\nu_{k,1}^{(j)},\dots,\nu_{k,N}^{(j)})$ for $j = 1, \dots, J$ as it has all the relevant intermediate values required for performing this operation.

\begin{definition}
The communication load $L$ of a scheme is the total amount of data (in bits) transmitted by the servers during the Shuffle phase normalized by $JQB$.
\end{definition}
\begin{example}
\label{ex:motivating}
Suppose that our task consists of $J=4$ jobs. For the $j$-th job we need to count $Q=6$ words given by the set $\mathcal{A}^{(j)}=\{\chi_1^{(j)},\dots,\chi_6^{(j)}\}$ in a book consisting of $N=6$ chapters using a cluster of $K=6$ servers.
$\mathcal{J}_j$ is associated with the $j$-th book and its subfiles with the chapters $1^{(j)},\dots,6^{(j)}$. Function $\phi_k^{(j)},k=1,\dots,Q$  (assigned to server $U_k$ since $Q=K$ as discussed) counts the word $\chi_k^{(j)}$ of $\mathcal{A}^{(j)}$ in the book indexed with $j$. 
This formulation fits the linear aggregation case precisely. Indeed, each reducer only needs the sum of the word counts for the subfiles that it does not locally store and hence there is scope for ``compressing" multiple values at the end of the Map phase.
\end{example}

\section{Description of the CAMR Scheme}
\label{sec:protocol}
In this section, we describe our proposed algorithm. We begin by introducing a few design theory definitions. 

\begin{definition}
A \emph{design} is a pair $(\mathcal{X}, \mathcal{A})$ consisting of
\begin{enumerate}
\item a set of elements (\emph{points}), $\mathcal{X}$, and
\item a family $\mathcal{A}$ (i.e. multiset) of nonempty subsets of $\mathcal{X}$ called \emph{blocks}, where each block has the same cardinality.
\end{enumerate}
\end{definition}

In this paper, we use a special class of designs, called \emph{resolvable designs}.

\begin{definition}
A subset $\mathcal{P}\subset \mathcal{A}$ in a design $(\mathcal{X},\mathcal{A})$ is said to be a \textit{parallel class} if for $X_i\in\mathcal{P}$ and  $X_j\in\mathcal{P}$ with $i\neq j$ we have $X_i\cap X_j=\emptyset$  and $\cup_{\{j:X_j\in P\}}X_j=\mathcal{X}$. A partition of $\mathcal{A}$ into several parallel classes is called a resolution, and $(\mathcal{X},\mathcal{A})$ is said to be a resolvable design if $\mathcal{A}$ has at least one resolution.
\end{definition}

It turns out that there is a systematic procedure for constructing resolvable designs from error correcting codes. 

Let $\mathbb{Z}_q$ denote the additive group of integers modulo $q$. The generator matrix of an $(k,k-1)$ single parity-check (SPC) code over $\mathbb{Z}_q$\footnote{We emphasize that this construction works even if $q$ is not a prime, i.e., $\mathbb{Z}_q$ is not a field.} is defined by
\begin{equation*}
\mathbf{G}_{SPC}=
\begin{bmatrix}
& &\vline&1\\
&\huge \mathbf{I}_{k-1}&\vline&\vdots\\
&&\vline&1
\end{bmatrix}.
\end{equation*}
This code has $q^{k-1}$ codewords. The codewords are $\mathbf{c} = \mathbf{u} \cdot \mathbf{G}_{SPC}$ for each possible message vector $\mathbf{u}$. 
The $q^{k-1}$ codewords $\mathbf{c}_i$ computed in this manner are stacked into the columns of a matrix $\mathbf{T}$ of size $k\times q^{k-1}$, i.e., 
\begin{equation*}
\mathbf{T}=[{\mathbf{c}}_1^T,{\mathbf{c}}_2^T,\cdots,{\mathbf{c}}_{q^{k-1}}^T].
\end{equation*}
The corresponding resolvable design is constructed as follows. Let $\mathcal{X}_{SPC} = [q^{k-1}]$ (for a positive integer $n$, we use $[n]$ to denote the set $\{1,2,\dots,n\}$ throughout) represent the point set of the design.
We define the blocks as follows. For $0 \leq l \leq q-1$, let $B_{i,l}$ be a block defined as
\begin{equation}
\label{eq:block_desc}
B_{i,l}=\{j: \mathbf{T}_{i,j}=l\}.
\end{equation}
The set of blocks $\mathcal{A}_{SPC}$ is given by the collection of all $B_{i,l}$ for $1 \leq i \leq k$ and $0 \leq l \leq q-1$ so that $|\mathcal{A}_{SPC}| = kq$. The following lemma (see  \cite{TangR18} for a proof in a different context) shows that this construction yields a resolvable design.

\theoremstyle{lemma}
\begin{lemma}
\label{lemma:cons_yields_resolv_design}
The above scheme always yields a resolvable design $(\mathcal{X}_{SPC},\mathcal{A}_{SPC})$ with $\mathcal{X}_{SPC}=[q^{k-1}]$, $|B_{i,l}|=q^{k-2}$ for all $1\leq i\leq k$ and $0\leq l\leq q-1$. The parallel classes are analytically described by $\mathcal{P}_i=\{B_{i,l}:0\leq l\leq q-1\}$, for $1\leq i\leq k$.
\end{lemma}

\DecMargin{1em}
\begin{algorithm}[!t]
\LinesNotNumbered
\label{alg:placement}
\KwIn{$J$ jobs, owner sets $\{X^{(j)}, j=1,\dots,J\}$, $k$ used in SPC code construction, batch size $\gamma$.}
Set $N=k\gamma$.\\
\For{each job $\mathcal{J}_j$}{
Split the data set of $\mathcal{J}_j$ into $N$ disjoint subfiles
\nonl
{
\abovedisplayskip=0pt
\belowdisplayskip=0pt
$$\{1^{(j)},\dots,N^{(j)}\}$$\\
\nonl and partition them into $k$ batches of $\gamma$ subfiles each.\\
Let $X^{(j)}=\{U_{i_1},\dots,U_{i_k}\}$. Label each batch with\\
a distinct index of an owner so that the batches are
$$\mathcal{B}=\{\mathcal{B}_{[i_1]}^{(j)},\dots,\mathcal{B}_{[i_k]}^{(j)}\}$$\\
\For{each owner $U_{k'}\in X^{(j)}$}{
Store all batches in $\mathcal{B}$ except $\mathcal{B}_{[i_{k'}]}^{(j)}$ in server $U_{k'}$.
}
}
}
\caption{File placement}
\end{algorithm}
\IncMargin{1em}

\subsection{Job assignment and file placement}
Our cluster consists of $K$ servers and we choose appropriate integers $q, k$ that factorize it as $K=k\times q$; we further need $N$ to be divisible by $k$. Next, we form a $(k, k-1)$ SPC code and the corresponding resolvable design, as described above. The jobs to be executed are associated with the point set $\mathcal{X}=[q^{k-1}]$. Hence $J=q^{k-1}$ and the block set $\mathcal{A}$ will be such that $|\mathcal{A}|=k\times q$. The servers are associated with the blocks and are indexed as $B_{i,j}, i=1,\dots,k,$ and $j=0,1,\dots,q-1$.

The assignment of jobs to servers follows the natural incidence between points and blocks. Thus, job $\mathcal{J}_j$ is processed by (or ``owned" by) the server indexed by $B_{i,l}$ if $j \in B_{i,l}$. For the sake of convenience we will also interchangeably work with servers indexed as $U_{1}, \dots, U_{K}$ with the implicit understanding that each $U_i, i \in [K]$ corresponds to a block from $\mathcal{A}$. By convention, server $U_i$ will be associated with the block $B_{\left \lceil{i/q}\right \rceil,\ (i-1)\ \mathrm{mod}\ q}$.

Let us denote the owners of $\mathcal{J}_j$ by $X^{(j)} \subset \{U_1, \dots, U_K\}$.
For each job, the data set is split into $k$ batches and each batch is made up of $\gamma$ subfiles, for some integer $\gamma>1$ (recall that $k|N$). The file placement policy is illustrated in Algorithm \ref{alg:placement}.

Each server is owner of $q^{k-2}$ jobs (block size). For each such job it participates in $k-1$ batches of size $\gamma$, as explained in Algorithm \ref{alg:placement}. Our requirement for the storage fraction is
$$\mu=\frac{q^{k-2}\cdot(k-1)\cdot\gamma}{Jk\gamma}=\frac{k-1}{K}.$$

\begin{example}
In Example \ref{ex:motivating}, we have a cluster of $K=6$ nodes. We chose our parameters $q=2$ and $k=3$, then we need to execute $J=q^{k-1}=4$ MapReduce jobs. The codewords for this choice of parameters are $\{000,011,101,110\}$. Hence, based on Eq. \eqref{eq:block_desc}, the owners %
are 

\begin{equation}
\label{eq:ex_owners}
\begin{array}{r@{}l}
X^{(1)}&{}=\{U_1, U_3, U_5\},\quad X^{(2)}=\{U_1, U_4, U_6\},\\
X^{(3)}&{}=\{U_2, U_3, U_6\},\quad X^{(4)}=\{U_2, U_4, U_5\}.
\end{array}
\end{equation}

\begin{figure}[t]
\centering
\includegraphics[scale=0.17]{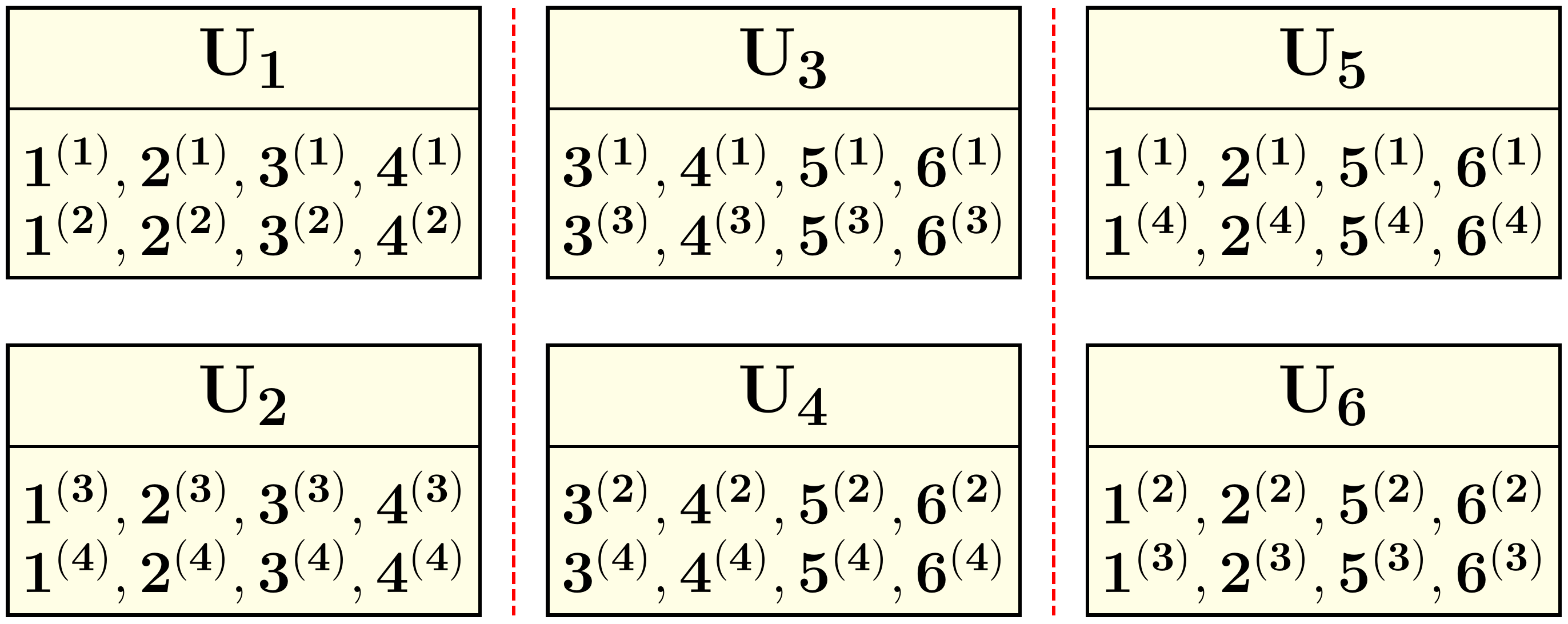}
\caption{Proposed placement scheme for $K=6$ servers and $N=6$ subfiles per computing job for $J=4$ jobs. The dotted lines show the partition of the servers into parallel classes.}
\label{fig:placement_ex1}
\end{figure}

We have subdivided the original data set of each job into $N=6$ subfiles. 
The subfiles of the $j$-th job are partitioned into three batches, namely $\{1^{(j)},2^{(j)}\}$, $\{3^{(j)},4^{(j)}\}$ and $\{5^{(j)},6^{(j)}\}$. Exactly four such batches are stored on each machine (\emph{cf.} Fig. \ref{fig:placement_ex1}). For $\gamma=2$, each job's data set is split into $N=k\gamma=6$ subfiles placed on a unique subset of $k=3$ nodes. For example, the subfiles of job $\mathcal{J}_1$, $\{1^{(1)},2^{(1)},\dots,6^{(1)}\}$, are stored exclusively on $U_1$, $U_3$ and $U_5$. Specifically, the three batches of the first job are
\begin{eqnarray*}
\mathcal{B}_{[i_3]}^{(1)}=\{1^{(1)}, 2^{(1)}\},\ \mathcal{B}_{[i_5]}^{(1)}=\{3^{(1)}, 4^{(1)}\},\ \mathcal{B}_{[i_1]}^{(1)}=\{5^{(1)}, 6^{(1)}\}.
\end{eqnarray*}

Then, batch $\mathcal{B}_{[i_3]}^{(1)}$ is stored on machines $U_1$ and $U_5$, $\mathcal{B}_{[i_5]}^{(1)}$ on $U_1$ and $U_3$ and, finally, $\mathcal{B}_{[i_1]}^{(1)}$ on $U_3$ and $U_5$. 
Each machine locally stores $\mu=\frac{k-1}{K}=\frac{1}{3}$ of all the data sets.
\end{example}

\subsection{Map phase}
During this phase, each server maps all the subfiles of each job it has partially stored, for all output functions. The resulting intermediate values have the form
$$\nu_{q,n}^{(j)}=\phi_q^{(j)}(n^{(j)}),\quad q\in[Q],\ n\in[N],\ j\in[J].$$

At the end of the Map phase, 
for each job $\mathcal{J}_j$, each mapper combines all those values $\nu_{q,n}^{(j)}$ that are indexed with  the same $q$ and $j$ (in other words, associated with the same function and job) and belong to the same batch of subfiles; we have already referred to this operation as aggregation. Our shuffle algorithm operates on the batch-level, as it will become clear in the following section.

\subsection{Shuffle phase}
\label{sec:shuffle_scheme}
The CAMR scheme carries out the data shuffling phase in three stages. The first two stages 
utilize a common shuffling algorithm (\emph{cf}. Algorithm \ref{alg:shuffling_lemma}), summarized in the following lemma and proved in \cite[Appendix]{arxiv_full_ISIT}. 

\begin{lemma}
\label{lem:shufffling_lemma}
 Consider a group of $k$ machines $G=\{U_1,\dots,U_k\}$ with the property that every subset of $G$ of the form $G\setminus\{U_{k'}\}$, stores a chunk of data of size $B$ bits, denoted $\mathcal{D}_{[k']}$, that $U_{k'}$ does not store. Then, there exists a protocol where each machine in $G$ can multicast a coded packet useful to all other $k-1$ machines and after $k$ such transmissions each of them can recover its missing chunk. The total number of bits transmitted in this protocol is $Bk/(k-1)$.
\end{lemma}

\begin{algorithm}[!t]
\LinesNotNumbered
\label{alg:shuffling_lemma}
\KwIn{Group of machines $G=\{U_1,\dots,U_k\}$,\\
data chunks $\{\mathcal{D}_{[k']}: U_{k'}\in G\}$.}
{
\abovedisplayskip=0pt
\belowdisplayskip=0pt
\For{each chunk $\mathcal{D}_{[k']}$}{
Split the chunk into $k-1$ disjoint packets
$$\{\mathcal{D}_{[k']}[i]:i=1,\dots,k-1\}$$\\
Let subset $G\setminus\{U_{k'}\}=\{U_1^{[k']},\dots,U_{k-1}^{[k']}\}$.\\
\For{each $i=1,\dots,k-1$}{
Associate packet $\mathcal{D}_{[k']}[i]$ with machine $U_i^{[k']}$.
}
}
\For{each machine $U_m\in G$}{
$U_m$ broadcasts\footnotemark
\begin{equation}
\label{eq:lemma_broadcast}
\Delta_m=\underset{k': U_m\in G\setminus\{U_{k'}\}}{\oplus}\{\mathcal{D}_{[k']}[i]: U_m=U_i^{[k']}\}
\end{equation}
}
}
\caption{Shuffling algorithm of Lemma \ref{lem:shufffling_lemma}}
\end{algorithm}

\begin{enumerate}[wide, labelwidth=!, labelindent=0pt]
\item \textbf{Stage 1}: In this stage, the owners of each job communicate among themselves. Let us fix a job $\mathcal{J}_j$ and consider the servers in $X^{(j)}\setminus \{U_{k'}\}$ of cardinality $k-1$ (\emph{cf.} Algorithm \ref{alg:placement}). During the Map phase, each machine in that subset has computed 
an aggregate needed by the remaining owner $U_{k'}$ which is
$$\alpha_{[k']}^{(j)}=\alpha(\{\nu_{k',n}^{(j)}:\ n\in \mathcal{B}_{[i_{k'}]}^{(j)}\}).$$

Repeating this process for every value of $j$ and $k'$, we can identify all aggregates $\alpha_{[k']}^{(j)}$. We shall now see an one-to-one correspondence between this setup and Lemma \ref{lem:shufffling_lemma} which is the following 
$$G=X^{(j)}\quad \text{and}\quad \mathcal{D}_{[k']}=\alpha_{[k']}^{(j)}$$
for $j=1,\dots,J$ and the owners $\{U_{k'}\in X^{(j)}\}$.

Each owner of a particular job, after receiving $k-1$ such values (one from every other owner of a particular job), can decode all of its missing aggregates for that job. 

\begin{example}
In Example \ref{ex:motivating}, let us consider the group of servers $\{U_1,U_3,U_5\}$ which are the owners of $\mathcal{J}_1$, storing $\{1^{(1)},2^{(1)},3^{(1)},4^{(1)}\}$, $\{3^{(1)},4^{(1)},5^{(1)},6^{(1)}\}$ and $\{1^{(1)},2^{(1)},5^{(1)},6^{(1)}\}$, respectively. Based on this allocation policy, server $U_1$ needs $\phi_1^{(1)}$ evaluations of the batch $\{5^{(1)}, 6^{(1)}\}$, i.e.,
$\nu_{1,5}^{(1)} \text{ and } \nu_{1,6}^{(1)}$ for $\mathcal{J}_1$ or simply the aggregate
$$\alpha(\nu_{1,5}^{(1)}, \nu_{1,6}^{(1)})=\nu_{1,5}^{(1)}+\nu_{1,6}^{(1)}.$$
Similarly, $U_3$ needs $\alpha(\nu_{3,1}^{(1)}, \nu_{3,2}^{(1)})$ and $U_5$ needs $\alpha(\nu_{5,3}^{(1)}, \nu_{5,4}^{(1)})$.

Next, we refer to Fig. \ref{fig:shuffle_ex1}. 
The compressed intermediate values are represented by circle/green, star/blue and triangle/red. We further suppose that each value can be split into two packets (represented by the left and right parts of each shape). If $U_1$ transmits left circle XOR left star, then $U_3$ is able to cancel out the star part (since $U_3$ also maps $\{3^{(1)}, 4^{(1)}\}$) and recover the circle part. Similarly, $U_5$ can recover the star part from the same transmission. Each of these transmissions is useful to two servers.

We can repeat this process for the remaining jobs. 
The total number of bits transmitted in this case is therefore $J\times k\times B/2=6B$. The incurred communication load is
$L_{\text{stage 1}}=\frac{6B}{JQB}=\frac{1}{4}$.
\end{example}

\footnotetext{The operation Eq. \eqref{eq:lemma_broadcast} is a bitwise XOR.}

\item \textbf{Stage 2}: In this stage, we form communication groups of both owners and non-owners of a job, so that the latter can recover appropriate data to reduce their functions.

Towards this end, we form collections of user groups by choosing one block from each parallel class based on a simple rule. We choose servers $B_{1,j_1}, B_{2, j_2}, \dots, B_{k,j_k}$ such that $\cap_{\ell=1}^k B_{\ell, j_\ell} = \emptyset$.
It has been proved in \cite{TangR18} (but in a different context) that if we remove a server $U_{k'}$  from such a group $G$, the servers in the corresponding subset $P=G\setminus \{U_{k'}\}$ of cardinality $|P|=k-1$ jointly own a job, say $\mathcal{J}_j$, that the remaining server $k'$ does not. In addition, based on the file placement policy described before (\emph{cf.} Algorithm \ref{alg:placement}), they share the batch of subfiles $\mathcal{B}_{[i_l]}^{(j)}$ for that common job and some $U_l\in X^{(j)}$. 

The following simple observation is important. By construction, $U_l$ is precisely the remaining owner of $\mathcal{J}_j$ and it should lie in the parallel class that none of the other owners belong to; that is the same class as of $U_{k'}$.  

During the Map phase, each node in $P$ has computed 
an aggregate needed by $U_{k'}$ which is
\begin{equation}
\label{eq:aggreagates_stage2}
\beta_{[k']}^{(j)}=\alpha(\{\nu_{k',n}^{(j)}:\ n\in \mathcal{B}_{[i_l]}^{(j)}\}).
\end{equation}

\begin{figure}[t]
\centering
\includegraphics[scale=0.62]{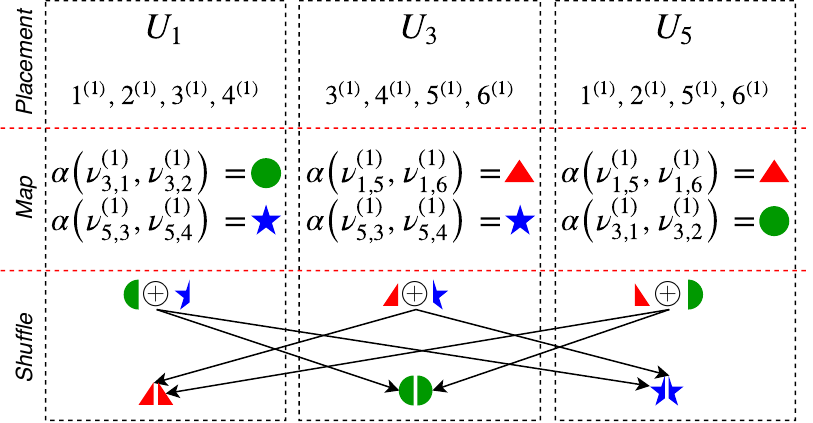}
\caption{Coded multicast transmission among the owners of $\mathcal{J}_1$ during stage 1.}
\label{fig:shuffle_ex1}
\end{figure}

As in stage 1, Lemma \ref{lem:shufffling_lemma} fits in this description and Algorithm \ref{alg:shuffling_lemma} defines the communication scheme; the shuffling group is $G$ and each server $U_{k'}\in G$ needs to recover the chunk $\mathcal{D}_{[k']}=\beta_{[k']}^{(j)}$ for the unique batch that all nodes in $P$ share.

As a result, at the end of stage 2, each server $U_{k'}, k'\in[K]$ is able to decode all aggregates of the form in Eq. \eqref{eq:aggreagates_stage2} for all values of $l$, i.e., for all nodes $U_l$ that belong to the same parallel class as $U_{k'}$. Note that each such value (for a fixed $l$) corresponds to $q^{k-2}$ (block size) jobs for which $U_{k'}$ does not store any subfiles and $U_l$ does not store the batch $\mathcal{B}_{[i_l]}^{(j)}$. 

\begin{example}
In Example \ref{ex:motivating}, in stage 2, the nodes recover values of jobs for which they haven't stored any subfile. Let $G=\{U_1,U_3,U_6\}$. Observe from Eq. \eqref{eq:ex_owners} that there is no job common to all three but each subset of two of them shares a batch of a job they commonly own. The remaining server needs an aggregate value of those subfiles. The values that each of $U_1,U_3,U_6$ needs as well as the corresponding transmissions are illustrated in Table \ref{table:ex_stage2}. We denote the $i$-th packet of an aggregate value by $\alpha(\cdot)[i]$.

There are $q^{(k-1)}(q-1)$ possible such groups we can pick. The total load is $L_{\text{stage 2}}=\frac{4\times3\times B/2}{JQB}=\frac{6B}{JQB}=\frac{1}{4}$.
\end{example}

\item \textbf{Stage 3}: Each worker is still missing values for jobs that it is not owner of from Stage 2. Now, servers communicate within parallel classes. In particular, we show in \cite[Appendix]{arxiv_full_ISIT} that all values that a server $U_m$ still needs can be aggregated and transmitted by a single owner-server in the same parallel class that $U_m$ belongs to. This server is unique and transmits one aggregate value of its jobs to every other server in the same parallel class.

Recall that the $i$-th class is $\mathcal{P}_i=\{B_{i,j},j=0,\dots,q-1\}$, then, server $U_k\in \mathcal{P}_i$ transmits
\begin{equation}
\label{eq:shuffle_stage3}
\Delta_k^{\text{stage 3}}=\alpha\Bigg(\bigcup\limits_{l:U_l\in X^{(j)}\setminus \{U_k\}}\{\nu_{m,n}^{(j)}:\ n\in \mathcal{B}_{[i_l]}^{(j)}\}\Bigg).
\end{equation}
to another $U_m\in\mathcal{P}_i$; obviously, $U_m\notin X^{(j)}$.

We repeat this process for every pair $(U_m,U_k)$ of servers in the same class.

\begin{example}
In Example \ref{ex:motivating}, if we consider the same group as in Stage 2, i.e., $G=\{U_1,U_3,U_6\}$ then we can see that $U_1$ still misses values $\nu_{1,1}^{(3)}, \nu_{1,2}^{(3)}, \nu_{1,3}^{(3)}$ and $\nu_{1,4}^{(3)}$ of $\mathcal{J}_3$ or simply their aggregate $\alpha(\nu_{1,1}^{(3)}, \nu_{1,2}^{(3)}, \nu_{1,3}^{(3)}, \nu_{1,4}^{(3)})$. Observe that all required subfiles locally reside in the cache of $U_2$ which can transmit the value to $U_1$. For the complete set of unicast transmissions see \cite[Table \ref{table:ex_stage2_missing}]{arxiv_full_ISIT}.
The load turns out to be $L_{\text{stage 3}}=\frac{6\times2\times B}{JQB}=\frac{1}{2}$.

The communication load of all phases is then $L_{\text{CAMR}}=1$.
Similarly, the load achieved by the CCDC scheme of \cite{compressed_CDC} for the same storage fraction $\mu=1/3$ is $L_{\text{CCDC}}=1$. Nonetheless, their approach would require a minimum of $J={{6}\choose{3}}=20$ distributed jobs to be executed.
\end{example}
\end{enumerate}

\begin{table}[t]
\centering
\caption{Transmissions within group of $\{U_1,U_3,U_6\}$ during stage 2}
\label{table:ex_stage2}
\begin{tabular}{|P{0.7cm}||P{4.1cm}|P{1.6cm}|}
\hline
Server&Transmits&Recovers\\
\hline
$U_1$&$\alpha(\nu_{6,3}^{(1)}, \nu_{6,4}^{(1)})[1]\oplus\alpha(\nu_{3,1}^{(2)}, \nu_{3,2}^{(2)})[1]$&$\alpha(\nu_{1,5}^{(3)}, \nu_{1,6}^{(3)})$\\
\hline
$U_3$&$\alpha(\nu_{6,3}^{(1)}, \nu_{6,4}^{(1)})[2]\oplus\alpha(\nu_{1,5}^{(3)}, \nu_{1,6}^{(3)})[1]$&$\alpha(\nu_{3,1}^{(2)}, \nu_{3,2}^{(2)})$\\
\hline
$U_6$&$\alpha(\nu_{3,1}^{(2)}, \nu_{3,2}^{(2)})[2]\oplus\alpha(\nu_{1,5}^{(3)}, \nu_{1,6}^{(3)})[2]$&$\alpha(\nu_{6,3}^{(1)}, \nu_{6,4}^{(1)})$\\
\hline
\end{tabular}
\end{table}

\subsection{Reduce phase}
Using the values it has computed and received, $U_k$ reduces
$$\phi_k^{(j)}(1^{(j)},\dots,N^{(j)})=\alpha(\nu_{k,1}^{(j)},\nu_{k,2}^{(j)},\dots,\nu_{k,N}^{(j)})$$
for all $k=1,\dots,K$ and $j=1,\dots,J$.

\section{Communication Load Analysis}
\label{sec:load}
In the first stage, for each of the $J$ jobs, each of the $k$ owners computes one aggregate and is associated with a unique corresponding packet of it, of size $\frac{B}{k-1}$. As a result, the communication load exerted in this stage is
$$L_{\text{stage 1}}=\frac{Jk\frac{B}{k-1}}{JQB}=\frac{k}{K(k-1)}.$$

The second stage involves the communication within all possible $q^{k-1}(q-1)$ groups that satisfy the desired property. In each case, $k$ workers transmit one value each, and the transmission is of length $\frac{B}{k-1}$. Then,
$$L_{\text{stage 2}}=\frac{q^{k-1}(q-1)k\frac{B}{k-1}}{JQB}=\frac{(q-1)k}{K(k-1)}.$$

Each server does not own $J-q^{k-2}$ jobs. For each of them, during stage 3, one transmission (of length $B$) from a server in the same parallel class is sufficient. Thus,
$$L_{\text{stage 3}}=\frac{K\left(J-q^{k-2}\right)B}{JQB}=\frac{q-1}{q}.$$

The total load is
$$L_{\text{CAMR}}=\sum\limits_{i=1}^3L_{\text{stage i}}=\frac{k(q-1)+1}{q(k-1)}.$$

\section{Comparison With Other Schemes}
\label{sec:load_comparison}
The technique proposed in \cite{compressed_CDC} demonstrates a load of 
\begin{equation}
\label{eq:load_ccdc}
L_{\text{CCDC}}=\frac{(1-\mu)(\mu K+1)}{\mu K}.
\end{equation}
for a suitable storage fraction such that $\mu K \in\{1,\dots,K-1\}$. Our storage requirement is equal to $\mu=\frac{k-1}{K}$. For the same storage requirement, Eq. \eqref{eq:load_ccdc} yields
\begin{eqnarray*}
L_{\text{CCDC}}&=&\frac{(1-\frac{k-1}{K})(\frac{k-1}{K} K+1)}{\frac{k-1}{K} K}=\frac{k(q-1)+1}{q(k-1)}.
\end{eqnarray*}
We conclude that the loads induced by the two schemes are identical.
However, their approach fundamentally relies on the requirement that the minimum number of jobs to be executed is $J_{\text{CCDC,\ min}}={{K}\choose{\mu K+1}}$. Comparing this value with our requirement for $J_{\text{CAMR}}=q^{k-1}$ and using a known bound for the binomial coefficients, we deduce that \cite{CLRS_book} 


\begin{eqnarray*}
J_{\text{CCDC,\ min}}={{K}\choose{\mu K+1}}={{kq}\choose{k}}\labelrel\geq{eq:J_bound}\left(\frac{kq}{k}\right)^k
\labelrel>{eq:CAMR_bound}J_{\text{CAMR,\ min}},
\end{eqnarray*}
where the bound of \eqref{eq:J_bound} is maximum when $q=2$ and becomes stricter for $q>2$; however, as $q$ increases the bound of \eqref{eq:CAMR_bound} loosens and it turns out that our requirement for the number of jobs becomes exponentially smaller than that of CCDC (\emph{cf.} \cite[Table \ref{table:J_comparison}]{arxiv_full_ISIT} for a numerical comparison).

\bibliographystyle{IEEEtran}
\bibliography{./bib/citations}

\begin{thebibliography}{10}
\providecommand{\url}[1]{#1}
\csname url@samestyle\endcsname
\providecommand{\newblock}{\relax}
\providecommand{\bibinfo}[2]{#2}
\providecommand{\BIBentrySTDinterwordspacing}{\spaceskip=0pt\relax}
\providecommand{\BIBentryALTinterwordstretchfactor}{4}
\providecommand{\BIBentryALTinterwordspacing}{\spaceskip=\fontdimen2\font plus
\BIBentryALTinterwordstretchfactor\fontdimen3\font minus
  \fontdimen4\font\relax}
\providecommand{\BIBforeignlanguage}[2]{{%
\expandafter\ifx\csname l@#1\endcsname\relax
\typeout{** WARNING: IEEEtran.bst: No hyphenation pattern has been}%
\typeout{** loaded for the language `#1'. Using the pattern for}%
\typeout{** the default language instead.}%
\else
\language=\csname l@#1\endcsname
\fi
#2}}
\providecommand{\BIBdecl}{\relax}
\BIBdecl

\bibitem{DeanG08}
J.~Dean and S.~Ghemawat, ``Mapreduce: Simplified data processing on large
  clusters,'' \emph{Communications of the ACM}, vol.~51, no.~1, pp. 107--113,
  January 2008.

\bibitem{HADOOP}
\BIBentryALTinterwordspacing
``{Apache Hadoop}.'' [Online]. Available: \url{http://hadoop.apache.org/}
\BIBentrySTDinterwordspacing

\bibitem{apache_spark}
M.~Zaharia, M.~Chowdhury, M.~J. Franklin, S.~Shenker, and I.~Stoica, ``Spark:
  Cluster computing with working sets,'' in \emph{2nd USENIX Conference on Hot
  Topics in Cloud Computing}, June 2010, pp. 10--10.

\bibitem{compressed_CDC}
S.~Li, M.~A. Maddah{-}Ali, and A.~S. Avestimehr, ``Compressed coded distributed
  computing,'' in \emph{{IEEE} International Symposium on Information Theory
  ({ISIT})}, June 2018, pp. 2032--2036.

\bibitem{he_resnet}
K.~He, X.~Zhang, S.~Ren, and J.~Sun, ``Deep residual learning for image
  recognition,'' in \emph{IEEE Conference on Computer Vision and Pattern
  Recognition (CVPR)}, June 2016, pp. 770--778.

\bibitem{saurav_g_analytics}
S.~Prakash, A.~Reisizadeh, R.~Pedarsani, and A.~S. Avestimehr, ``Coded
  computing for distributed graph analytics,'' in \emph{{IEEE} International
  Symposium on Information Theory ({ISIT})}, June 2018, pp. 1221--1225.

\bibitem{konstantinidis_ramamoorthy_globecom}
K.~Konstantinidis and A.~Ramamoorthy, ``Leveraging coding techniques for
  speeding up distributed computing,'' in \emph{IEEE Global Communications
  Conference (GLOBECOM)}, December 2018.

\bibitem{Chowdhury_etal11}
M.~Chowdhury, M.~Zaharia, J.~Ma, M.~I. Jordan, and I.~Stoica, ``Managing data
  transfers in computer clusters with orchestra,'' \emph{ACM SIGCOMM Computer
  Communication Review}, vol.~41, no.~4, pp. 98--109, August 2011.

\bibitem{GuoRZ13}
Y.~Guo, J.~Rao, and X.~Zhou, ``ishuffle: Improving hadoop performance with
  shuffle-on-write,'' in \emph{10th International Conference on Autonomic
  Computing (ICAC)}, June 2013, pp. 107--117.

\bibitem{Chen_graph14}
R.~Chen, X.~Ding, P.~Wang, H.~Chen, B.~Zang, and H.~Guan, ``Computation and
  communication efficient graph processing with distributed immutable view,''
  in \emph{23rd International Symposium on High-performance Parallel and
  Distributed Computing (HPDC)}, June 2014, pp. 215--226.

\bibitem{tandon_gradient}
R.~Tandon, Q.~Lei, A.~G. Dimakis, and N.~Karampatziakis, ``Gradient coding:
  Avoiding stragglers in distributed learning,'' in \emph{34th International
  Conference on Machine Learning (ICML)}, vol.~70, August 2017, pp. 3368--3376.

\bibitem{LiMA16}
S.~Li, M.~A. Maddah-Ali, Q.~Yu, and A.~S. Avestimehr, ``A fundamental tradeoff
  between computation and communication in distributed computing,'' \emph{IEEE
  Transactions on Information Theory}, vol.~64, no.~1, pp. 109--128, January
  2018.

\bibitem{TangR18}
L.~Tang and A.~Ramamoorthy, ``Coded caching schemes with reduced
  subpacketization from linear block codes,'' \emph{IEEE Transactions on
  Information Theory}, vol.~64, no.~4, pp. 3099--3120, April 2018.

\bibitem{arxiv_full_ISIT}
\BIBentryALTinterwordspacing
K.~Konstantinidis and A.~Ramamoorthy, ``{CAMR: Coded Aggregated MapReduce},''
  2019. [Online]. Available:
  \url{https://www.ece.iastate.edu/adityar/publications/}
\BIBentrySTDinterwordspacing

\end{thebibliography}

\clearpage
\appendix


\begin{table}[t]
\centering
\caption{Needed aggregate values at the end of stage 2}
\label{table:ex_stage2_missing}
\begin{tabular}{|P{0.7cm}||P{6.5cm}|}
\hline
Server&Needs\\
\hline
$U_1$&$\alpha(\nu_{1,1}^{(3)}, \nu_{1,2}^{(3)}, \nu_{1,3}^{(3)}, \nu_{1,4}^{(3)})$ and $\alpha(\nu_{1,1}^{(4)}, \nu_{1,2}^{(4)}, \nu_{1,3}^{(4)}, \nu_{1,4}^{(4)})$\\
\hline
$U_2$&$\alpha(\nu_{2,1}^{(1)}, \nu_{2,2}^{(1)}, \nu_{2,3}^{(1)}, \nu_{2,4}^{(1)})$ and $\alpha(\nu_{2,1}^{(2)}, \nu_{2,2}^{(2)}, \nu_{2,3}^{(2)}, \nu_{2,4}^{(2)})$\\
\hline
$U_3$&$\alpha(\nu_{3,3}^{(2)}, \nu_{3,4}^{(2)}, \nu_{3,5}^{(2)}, \nu_{3,6}^{(2)})$ and $\alpha(\nu_{3,3}^{(4)}, \nu_{3,4}^{(4)}, \nu_{3,5}^{(4)}, \nu_{3,6}^{(4)})$\\
\hline
$U_4$&$\alpha(\nu_{4,3}^{(1)}, \nu_{4,4}^{(1)}, \nu_{4,5}^{(1)}, \nu_{4,6}^{(1)})$ and $\alpha(\nu_{4,3}^{(3)}, \nu_{4,4}^{(3)}, \nu_{4,5}^{(3)}, \nu_{4,6}^{(3)})$\\
\hline
$U_5$&$\alpha(\nu_{5,1}^{(2)}, \nu_{5,2}^{(2)}, \nu_{5,5}^{(2)}, \nu_{5,6}^{(2)})$ and $\alpha(\nu_{5,1}^{(3)}, \nu_{5,2}^{(3)}, \nu_{5,5}^{(3)}, \nu_{5,6}^{(3)})$\\
\hline
$U_6$&$\alpha(\nu_{6,1}^{(1)}, \nu_{6,2}^{(1)}, \nu_{6,5}^{(1)}, \nu_{6,6}^{(1)})$ and $\alpha(\nu_{6,1}^{(4)}, \nu_{6,2}^{(4)}, \nu_{6,5}^{(4)}, \nu_{6,6}^{(4)})$\\
\hline
\end{tabular}
\end{table}

\subsection*{Proof of Lemma \ref{lem:shufffling_lemma}}
We shall refer to Algorithm \ref{alg:shuffling_lemma} in order to show that each machine in $G$ can recover its missing data chunk. Fix a pair of machines $\{U_m,U_k\}\subset G$ and the packet $\Delta_m$ transmitted from $U_m$ to $U_k$. By canceling out all terms of $\Delta_m$ with $U_k\in G\setminus\{U_{k'}\}$ in Eq. \eqref{eq:lemma_broadcast}, which $U_k$ locally stores, it can recover the remaining term, i.e., $\{\mathcal{D}_{[k]}[i]:U_m=U_i^{[k]}\}$. Keeping $U_{k}$ fixed, we repeat this process for every possible machine $U_m\in G\setminus\{U_{k}\}$. Since each of them is associated with a distinct packet of $\mathcal{D}_{[k]}$ it follows that by receiving the $k-1$ packets
$$\{\Delta_m: U_m\in G\setminus\{U_{k}\}\},$$
$U_k$ can recover the following packets
$$\{\mathcal{D}_{[k]}[i]:U_i\in G\setminus\{U_{k}\}\}.$$
Subsequently, $U_{k}$ concatenates them in order to recover $\mathcal{D}_{[k]}$. Since this proof holds independently of the choice of $U_m$, we have shown that all machines can recover their missing chunks at the end of the transmissions.

Since each chunk is assumed to be of size $B$ bits and it was split into $k-1$ packets of size $B/(k-1)$, the total amount of transmitted data is $Bk/(k-1)$.

\subsection*{Proof of Shuffling Correctness of Stage 3}
The proof follows from stage 2 and by the resolvability property of our design. Let us fix a shuffling group of stage 2, say $G$, a subset $P=G\setminus \{U_m\}$ and focus on the excluded server $U_m$. The servers in $P$ share a batch of a job $\mathcal{J}_j$ whose values have transmitted to $U_m$. The remaining batches of $\mathcal{J}_j$ that $U_m$ still needs are locally stored at a single server (precisely the owner of the job) in the remaining parallel class, i.e., the parallel class of $U_m$ (cf. [Section III.A]). The fact that the design is resolvable makes that server unique, since no blocks within a parallel class can have common points (recall that points have one-to-one correspondence with the jobs). That node will transmit the uncoded aggregate to $U_m$. Such transmissions benefit a single machine.

\begin{table}[t]
\centering
\caption{Comparison of the minimum requirement on the number of jobs for CAMR and CCDC schemes using the same storage fraction. Cluster consists of $K=100$ servers.}
\label{table:J_comparison}
\begin{tabular}{|P{0.5cm}||P{1.5cm}|P{1.5cm}|}
\hline
\multirow{2}{*}{k}&\multicolumn{2}{c|}{Minimum $J$}\\
\cline{2-3}
&CAMR&CCDC\\
\hline
2&50&4950\\
\hline
4&15625&3921225\\
\hline
5&160000&75287520\\
\hline
\end{tabular}
\end{table}

\end{document}